# Manifesting pseudo-spin polarization of graphene with field emission image

Jingkun Chen, Zhibing Li[*] and Weiliang Wang

State Key Laboratory of Optoelectronic Materials and Technologies, School of Physics and Engineering, Sun Yat-sen University, Guangzhou, P.R. China 510275

**Abstract**

Coherent emission of electron from graphene in both electric and magnetic fields is studied. We obtain the emission wave function analytically. The emission current density is calculated. The structure of Landau levels is recognizable in the emission image. The emission pattern depends on the phase difference of two sub-lattices. We find that the pattern changes obviously with the gate voltage on the edge. It provides a way to manipulate the emission pattern.

**1. Introduction**

Graphene can be described by two-dimensional chiral fermions with the pseudo-spin associating with the bi-lattice structure [1]. The effects of pseudo-spin have been are observed in a number of experiments [2-4]. Recently, a photoemission experiment manifested the Berry phase of graphene [5]. In the field of quantum information, people are interested in manipulation of pseudo-spin polarization. The present paper will investigate the possibility of extracting pseudo-spin polarization by

---

mean of electric field assisted electron emission. The Landau levels that are created by a magnetic field perpendicular to the graphene plane are crucial for this purpose.

Several groups have demonstrated that graphene has promising field emission properties such as low threshold field and large emission current density [6-14]. Theoretical studies suggested that the characteristic current-field relation of two-dimensional emitter disobeys the Fowler-Nordheim law [15-17]. Recently, an experiment showed interference fringes of the emission current as a direct evidence for the coherent cold field emission (CCFE) of graphene [6]. Theoretically, a dragonfly-like pattern of CCFE was predicted for an ideal graphene field emitter in the absent of magnetic field [18]. The present paper extends the theory of [18] to include magnetic fields.

A profound effect of the magnetic field on the two-dimensional electron gas is the quantum Hall effect (QHE) that has been realized in graphene [19, 20] and can be observed even at the room temperature [19-21]. The QHE of graphene is substantially different from that of non-relativistic electron gases [22] by the zero-energy Landau level which is absent in the latter case. In the present text of CCFE, Landau levels near the Fermi level play an important role. In the bulk graphene all states of a given Landau level are degenerate. The edge effect lifts the degeneracy such that each Landau level splits into two bands [23,24,25,26]. The splitting becomes significant when the magnetic field is larger than Tesla. In this regime, CCFE images of various edge states can be distinguished from each other, as we will show.

Section 2 defines the model and gives the analytical expressions for the emission wave function and the emission current line-density. Section 3 presents the numerical results of emission patterns, following with discussions. The last section is the conclusion.

## 2. Theory

The model is an individual graphene mounted vertically on a planar cathode, with the free edge (emission edge) terminated by hydrogen atoms. Let the graphene lie on the *x-y* plane in the region of $-h < x < 0$, with the emission edge along the *y* axis. A uniform macroscopic electric field $F_0$ is applied on the graphene in the negative *x* direction by an anode that is far away from the graphene presumably. In a proper approximation [18], the potential barrier for the *p*-th eigenstate of graphene has the form $V(\vec{r}) = W_p - eF_0\sqrt{h(x+\sqrt{x^2+z^2})}$. The zero-field effective barrier height reads

$$W_p = W_0 - E_p + \frac{(\hbar k)^2}{2m} \tag{1}$$

Where $W_0$=4.32eV is the work function [25] and $E_p$ is the eigenenergy relative to the Fermi level of the neutral graphene (referred to as intrinsic Fermi level). The last term is the transversal kinetic energy with *k* the *y*-directional wave number. The zigzag and armchair edges terminated by hydrogen atoms are two edges often discussed in literatures. However it was argued that the zigzag edge terminated with hydrogen atoms has ignorable contribution to field emission because that the relevant states have large transversal kinetic energies thereby a too large $W_p$ [16]. Therefore, the present paper concentrates on the armchair edge.

We treat the CCFE as coherent decay of edge atomic orbitals (EAOs). Because the π-orbitals mainly attribute to the relevant electronic properties of graphene, we may assume that the EAOs have the form of the π-orbitals. The smooth matching condition determines a characteristic length $1/\tau_p = \hbar/\sqrt{2mW_p}$ for each eigenstate. The path-decomposition method provides a bridge that connects the emission wave function and quantum states of the emitter. The phase information is preserved in this approach while the exact atomic potential, which is usually un-known, is not inquired [18].

Because most electric field in graphene is screened, one can first find the eigenstates of graphene in the absent of electric field. Then the level bending takes account of the residual macroscopic electric field in graphene. The low energy states near the intrinsic Fermi level are located in the Dirac cones where the dispersion relation is linear. There are two independent Dirac cones, one of them focuses at the Dirac point $\mathbf{K}_+ = (K,0)$ and the other at $\mathbf{K}_- = (-K,0)$. Where $K = \frac{4\pi}{3\sqrt{3}a}$ with the bond length $a = 0.142$ nm. In the magnetic field along the $z$ direction, an eigenstate of a given $k$ is localized in the $x$ direction when the Landau gauge $\mathbf{A} = (0, Bx, 0)$ is chosen. Without losing generality, we assume $B > 0$. The probability amplitude along the $x$ direction is described by a pair of Weyl spinors [23], $\Psi_p^{(L)} = (\psi_{p,A}^{(L)}(x), \psi_{p,B}^{(L)}(x))^{\mathrm{T}}$ and $\Psi_p^{(R)} = (\psi_{p,B}^{(R)}(x), \psi_{p,A}^{(R)}(x))^{\mathrm{T}}$, where the superscript $L(R)$ is referred to the left-handed (right-handed) spinor that associates with $\mathbf{K}_-$ ($\mathbf{K}_+$) Dirac point, while the subscript A (B) is referred to the pseudo-spin component that associates with the A(B)-sublattice. The symbol T means matrix transposition. We omit the electron spin indexes because the Zeeman effect and spin-orbital coupling have been neglected. The eigenenergy relative to the intrinsic Fermi level can be written as $E_p = s\sqrt{\lambda_p}\varepsilon_0$, where the characteristic energy $\varepsilon_0 = \sqrt{2}\hbar v_F / l_B$ with $v_F \sim c/330$ the Fermi velocity and $l_B = \sqrt{\frac{\hbar}{eB}}$. The signature $s = 1(-1)$ is for the positive (negative) energy states. The dimensionless parameter $\lambda_p$ is determined by the boundary condition.

Due to the $y$-directional symmetry, the wave number $k$ is conserved. In the graphene region ($x < 0$), the effective potential for each spinor component is a parabolic potential that is minimized at $\tilde{x}_0 = -\xi_0 = -kl_B$. In the states of large positive $k$, electron is localized around $\tilde{x}_0$ that is the distance from the edge. Therefore the edge effect is negligible for these states and their $\lambda_p$ converge to non-negative integers $n$ ($n = 0,1,2,\cdots$). We will label the Landau levels with $n$. The wave functions of the Weyl spinors can be given in the parabolic cylinder function $D_\lambda$ as

$$\psi_{p,A}^{(R)}(x) = e^{-i\frac{s\pi}{4}} N_p D_{\lambda_p}(\sqrt{2}\xi) \tag{2}$$

$$\psi_{p,B}^{(R)}(x) = ise^{-i\phi_p - i\frac{s\pi}{4}} \sqrt{\lambda_p} N_p D_{\lambda_p - 1}(\sqrt{2}\xi) \tag{3}$$

$$\psi_{p,A}^{(L)}(x) = \eta e^{-i\frac{s\pi}{4}} \sqrt{\lambda_p} N_p D_{\lambda_p - 1}(\sqrt{2}\xi) \tag{4}$$

$$\psi_{p,B}^{(L)}(x) = i\eta s e^{-i\phi_p - i\frac{s\pi}{4}} N_p D_{\lambda_p}(\sqrt{2}\xi) \tag{5}$$

Where $\xi = \frac{x}{l_B} + kl_B$, $\eta = \pm 1$, and $\phi_p$ is the phase difference of two sub-lattices. The normalization factor $N_p = (4l_B \int_{-\infty}^{\xi_0} d\xi D_{\lambda_p}^2)^{-\frac{1}{2}}$. The boundary conditions for the armchair edge are $\psi_{p,A}^{(R)}(0) + \psi_{p,A}^{(L)}(0) = 0$ and $\psi_{p,B}^{(R)}(0) + \psi_{p,B}^{(L)}(0) = 0$. Note that the boundary conditions mix the left and right spinors but are invariant with $\phi_p$. Substitution of (2-5) in the boundary conditions yields $D_{\lambda_p}(\sqrt{2}\xi_0) = -\eta\sqrt{\lambda_p}D_{\lambda_p-1}(\sqrt{2}\xi_0)$, from which one can determine $\lambda_p$. Note that the Hamiltonian conserves the time-Y parity that is defined as the combination transformation of time reverse and the y-directional reflection. With the constant phase factor $e^{-i\frac{s\pi}{4}}$ introduced in (2-5), the time-Y parity is equal to $\eta$. Therefore, an eigenstate is specified by four quantum numbers $(n,s,k,\eta)$ that have been denoted by p in previous text. The zero-energy Landau level with $n = 0$ is special for the massless Weyl spinor.[26,26] It also slits into two levels in the edge regime, both have $\eta = -1$.

Tunnelling through the two-dimensional barrier $V(\vec{r})$ can be solved by the conformal transformation that maps the x-z plane projection of the cathode and graphene into a straight line. In the transformed plane, the outer turning point of the potential energy barrier is $\tilde{x}_1 = \frac{W_p}{eF_0}$. The emission wave function has been obtained by the path-decomposition method in [18]. In a large distance it has the asymptotic form

$$\varphi_p(\vec{r}) \sim \tilde{C}_p (\frac{\tilde{x}_1}{x})^{1/4} \chi(z) S_p^{\frac{5}{4}} e^{-\frac{b_0}{2}S_p - \frac{1}{4}(2-S_p+\sqrt{S_p^2+4})} \frac{e^{iky}}{\sqrt{L}} \qquad (6)$$

where $\tilde{C}_p$ is the edge probability amplitude given by the tight-binding theory, $L$ is the length of the edge. The dimensionless factor $S_p = \frac{4\tau_p \tilde{x}_1^2}{9h}$ and the constant $b_0 = 1 + \frac{4}{5\sqrt{3}}$. The z-directional distribution is given by the square-normalized amplitude $\chi_p(z) = \frac{c_1 \mu_p^{3/4}}{\sqrt{x}} \frac{z}{x} e^{-\frac{\mu_p}{2}(\frac{z}{x})^2}$ where $\mu_p = \frac{c_2 k_a^2 h}{\tau_p}$ with $k_a = \frac{1}{\hbar}\sqrt{2meF_0 x}$. The two numerical constants $c_1 \sim 1.2$ and $c_2 \sim 0.45$.

Let $\Delta$ be the edge energy shift with respect to the intrinsic Fermi level. A finite $\Delta$ may be originated from the band-bending as a result of the field penetration [16, 18].

Applying a gate voltage on the edge may also cause a finite Δ. With the wave function (6) on hand, it is straight forward to calculate the emission image on the screen. Let the screen plane and anode plane be perpendicular to the *x*-axis. Two planes coincide at $x_a$. We obtain the electric current density on the screen

$$j(y,z) = \sum_p \frac{c_0 e \tau_p}{2\pi a^2 t_a} \frac{|\tilde{C}_p|^2 |\chi(z)|^2}{e^{(E_p-\Delta)/(k_B T)}+1} S_p^{\frac{5}{2}} e^{-b_0 S_p - \frac{1}{2}(2-S_p+\sqrt{S_p^2+4})} \Bigg|_{k=\frac{my}{\hbar t_a}} \quad (7)$$

Where *T* is the temperature of graphene and the numerical constant $c_0 \approx 0.1$. In the near axis region and for large $x_a$, the *k* wave will reach the screen at $y = \frac{\hbar k}{m} t_a$ with the flying time $t_a = \sqrt{\frac{2m x_a}{eF_0}}$. Using the spinor amplitudes given in (2-6), we obtain the probability of electron in the *p*-state to appear in the edge carbon orbitals as

$$|\tilde{C}_p|^2 = \frac{3^{5/2} a^2}{2} |N_p D_{\lambda_p}(\sqrt{2}\xi_0)|^2 (1-\eta s \sin\phi_p) \quad (8)$$

The phase difference of two sub-lattices can be written as $\phi_p = ka + \phi_0$ with $\phi_0$ a constant phase shift. The energy levels are degenerate with $\phi_0$ in the present model. In principle $\phi_0$ depends on state-preparing process and may change by interactions. For instance, the spin-orbital interaction can lift the degeneracy. The appearance of $\phi_0$ in (8) represents the pseudo-spin polarization. Averaging over $\phi_0$ with equal weight will get rid of the $\phi_0$-dependence, leading to the un-polarized probability.

## 3. Results and discussions

In the following numerical calculations, we fix $x_a = 1$ mm, $h = L = 10$ μm, $F_0 = 20$ V/μm, and $B = 15$ Tesla. Figure 1 presents a typical emission image on the screen at the room temperature, with $\Delta = 0$ and $\phi_0 = 0$. Two main peaks separating in *z* direction are caused by interference of two lops of the π-orbital. Integration over *z* gives the *y*-directional current line-density, as shown by the solid black curve in Figure 2 for the temperature of liquid nitrogen (77 Kelvin). Coloured curves are partial current line-densities from individual Landau levels close to the Fermi level (see inset of the same figure). Each partial current line-density is coloured with the same colour of its corresponding Landau level. Solid (dashed) curves are for positive (negative) levels. The Landau levels exhibit in the emission image as the multi-peak structure along the *y*-direction.

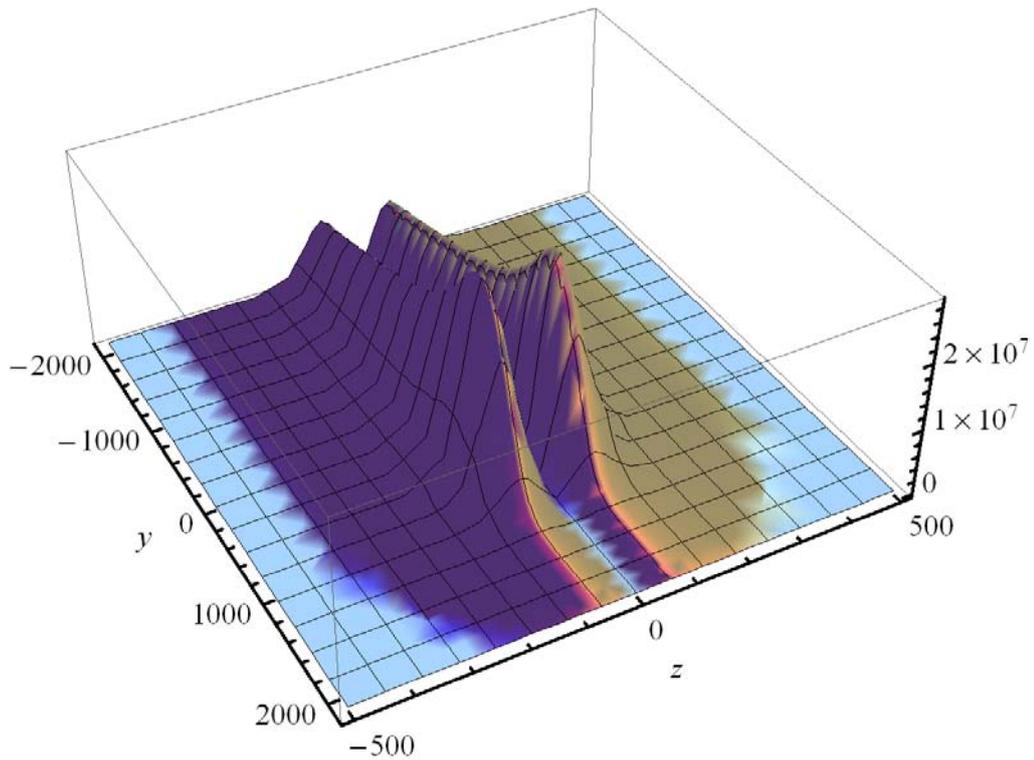

Figure 1 (colour on line) The emission pattern for $F_0 = 20$ V/μm and $B = 15$ Tesla, at the room temperature. The vertical axis is the emission current density on the screen (mA/cm$^2$). The *y* and *z* coordinates are in unit of nanometre. Both the edge energy shift and the phase shift are set to zero.

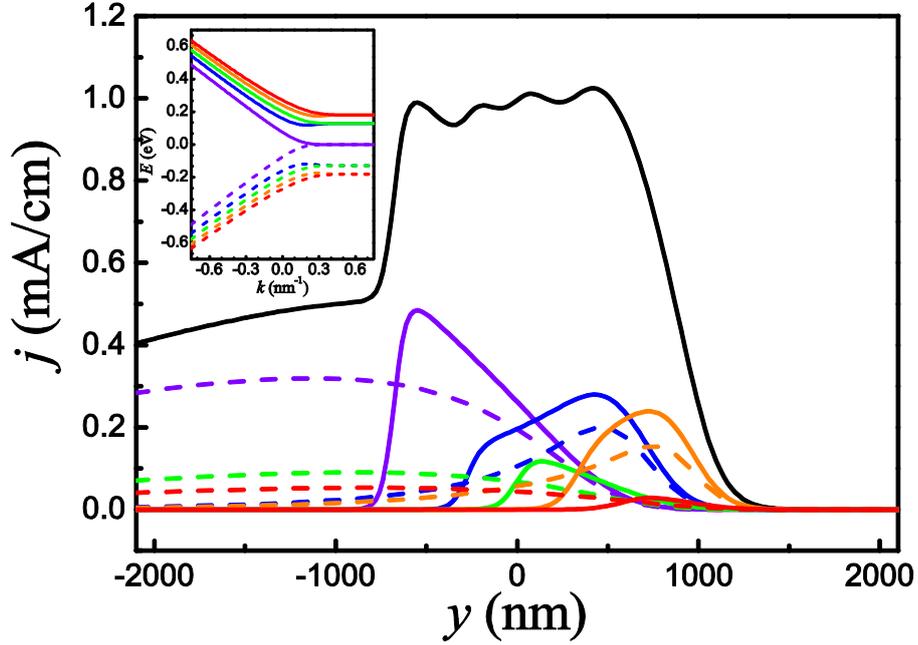

Figure 2 The *y*-directional current line-density on the screen for $F_0 = 20 \, \text{V}/\mu\text{m}$, $B = 15$ Tesla, $T = 77$ Kelvin, and $\Delta = 0.2$ eV. Black solid curve is the total line-density. Each coloured curve is the partial line-density corresponding to the Landau level as shown in the inset in the same colour.

To see the effect of the edge energy shift, the *y*-directional current line-densities on the screen for edge energy shifts from $\Delta = -0.2$ to $0.2$, increasing with step 0.05eV are plotted in Figure 3 (curves from the bottom to the top) at $T = 77$ Kelvin. The emission pattern depends on $\Delta$ obviously. The left wings of the current line-densities of positive Landau levels are sharply cut because the occupation numbers decrease exponentially with decreasing $k$ when the up-bending levels exceed $\Delta$. On the other hand, the negative Landau levels are fully occupied when $k$ is small enough because the levels are bending downwards with decreasing $k$. Therefore, the total line-densities of various edge energy shifts coincide with a smooth curve on the left side of Figure 3. The right wings of the line-densities are cut for two reasons: (1) depletion of the Landau levels that are above $\Delta$, (2) exponentially decreasing of the probability amplitude on the edge when $\tilde{x}_0$ moves to the inner region of graphene. Thus, the right wing of the current line-density is not sensitive to the temperature.

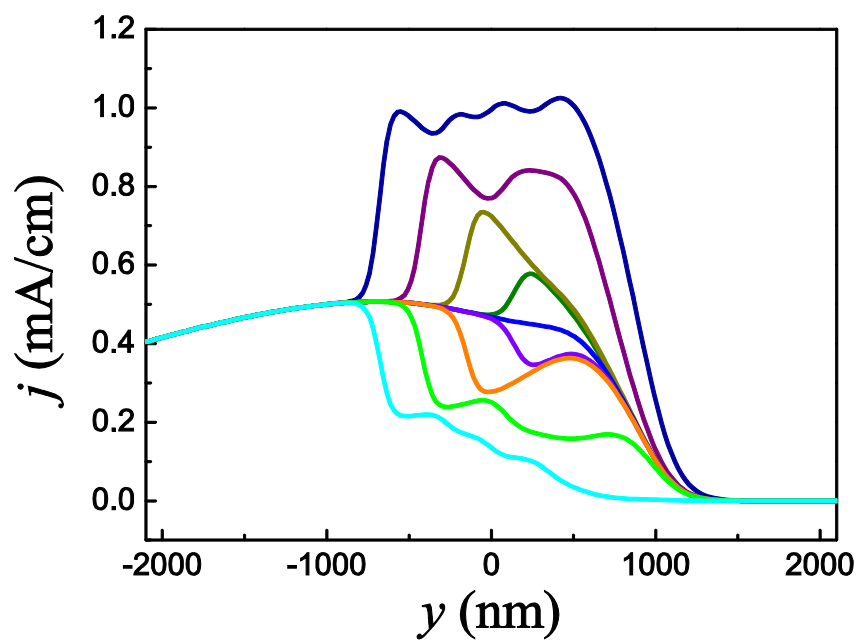

Figure 3 (colour on line) The *y*-directional current line-density on the screen for $F_0 = 20\,\text{V}/\mu\text{m}$, $B = 15\,\text{Tesla}$, and $T = 77\,\text{Kelvin}$. From the bottom to the top, the edge energy shifts range from $-0.2$ to $0.2$, increasing with step $0.05$ eV.

The effect of the pseudo-spin polarization is shown in Figure 4 for $\Delta = -0.2\,\text{eV}$ and $T = 77$ Kelvin. The patterns of $\phi_0 = 0$ and $\pi$ almost coincide with the un-polarized pattern (black dashed) both in the middle region and on the right wing. However, the left wings of them are clearly distinguishable. The patterns of $\phi_0 = \pm\frac{\pi}{2}$ are dramatically different from those of $\phi_0 = 0$ and $\pi$. The pattern of $\phi_0 = \frac{\pi}{2}$ has the main peak on the left and the second peak on the right. Its left wing is the lowest. On the other hand, the pattern of $\phi_0 = -\frac{\pi}{2}$ has the main peak on the right, the second peak on the left, and a fat tail on the left wing.

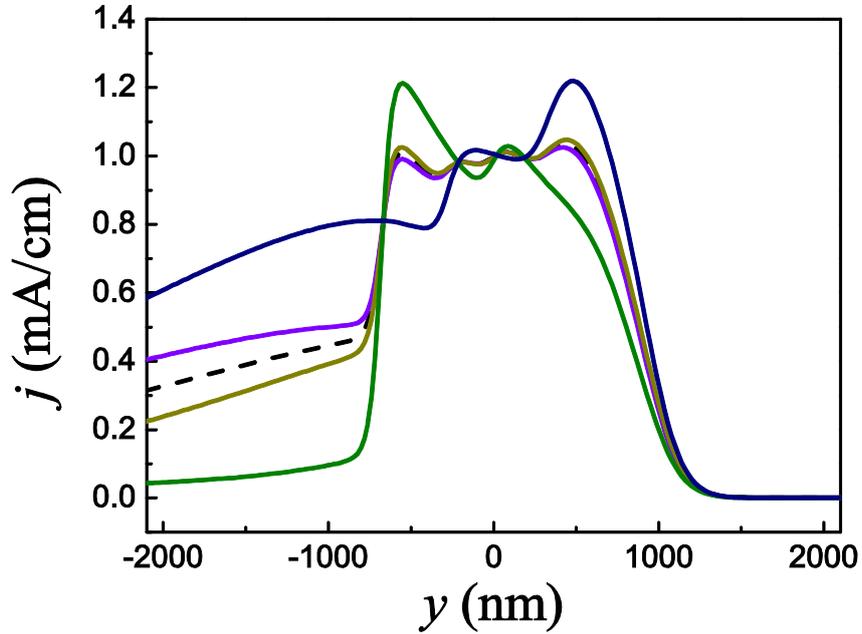

Figure 4 (colour on line) The phase dependence of the current line-density on the screen for $F_0 = 20\,\text{V}/\mu\text{m}$, $B = 15\,\text{Tesla}$, $T = 77\,\text{Kelvin}$, and $\Delta = 0.2\,\text{eV}$. The black dashed curve is the un-polarized line-density. The solid curves with their left wings from higher to lower are corresponding to phase shifts $-\pi/2, 0, \pi$, and $\pi/2$.

Incidentally, states with $\phi_0 = \frac{\pi}{2}$ or $\phi_0 = -\frac{\pi}{2}$ have the reflection symmetry on the *x-y* plane in the absent of fields. Therefore, one may build states of $\phi_0 = \pm\frac{\pi}{2}$ from states that possess the reflection symmetry.

## 4. Conclusions

We obtained the analytical expression for the coherent field emission current density of the armchair edge of graphene in both electric and magnetic fields. The emission image presents two interference fringes that can be traced back to the wave function of the π-orbital. The emission pattern manifests the structure of Landau levels. One may manipulate the pattern with a gate voltage on the edge. We discussed the effect of pseudo-spin polarization. The emission pattern changes dramatically when the phase difference of two sub-lattices changes by π/2. The relation between the emission pattern and the phase would enable one to read the pseudo-spin polarization from the coherent field emission image.

**Acknowledgements**

The project is supported by the National Basic Research Program of China (2013CB933601, 2007CB935500), the National Natural Science Foundation of China (11274393 and 11104358), and the Fundamental Research Funds for the Central Universities (No. 13lgpy34).

**Captions of figures:**

Figure 1 (colour on line) The emission pattern for $F_0 = 20\,\text{V}/\mu\text{m}$ and $B = 15\,\text{Tesla}$, at the room temperature. The vertical axis is the emission current density on the screen (mA/cm$^2$). The $y$ and $z$ coordinates are in unit of nanometre. Both the edge energy shift and the phase shift are set to zero.

Figure 2 The $y$-directional current line-density on the screen for $F_0 = 20\,\text{V}/\mu\text{m}$, $B = 15\,\text{Tesla}$, $T = 77\,\text{Kelvin}$, and $\Delta = 0.2$ eV. Black solid curve is the total line-density. Each coloured curve is the partial line-density corresponding to the Landau level as shown in the inset in the same colour.

Figure 3 (colour on line) The $y$-directional current line-density on the screen for $F_0 = 20\,\text{V}/\mu\text{m}$, $B = 15\,\text{Tesla}$, and $T = 77\,\text{Kelvin}$. From the bottom to the top, the edge energy shifts range from $-0.2$ to $0.2$, increasing with step $0.05$ eV.

Figure 4 (colour on line) The phase dependence of the current line-density on the screen for $F_0 = 20 \text{ V/µm}$, $B = 15 \text{ Tesla}$, $T = 77 \text{ Kelvin}$, and $\Delta = 0.2 \text{ eV}$. The black dashed curve is the un-polarized line-density. The solid curves with their left wings from higher to lower are corresponding to phase shifts $-\pi/2, 0, \pi$, and $\pi/2$.